\documentclass{cpbtex}

\usepackage{tabularx,graphicx, float, setspace, enumitem,  booktabs}
\usepackage{caption}
\usepackage{amsmath,amsfonts,amssymb,bm,upgreek}
\usepackage[compress,nospace]{cite}
\usepackage[dvipsnames]{xcolor}
\usepackage{natbib}
\usepackage{indentfirst}
\setcitestyle{numbers,square}

\begin{document}
\begin{CJK*}{UTF8}{song}

\title{Single-flux-quantum-based Qubit Control with Tunable Driving Strength}

% Title should be concise; avoid abbreviations if possible; and not begin with `A', `An', `The', or `Study on'.

\author{Kuang Liu $^{1,2,3}$, Yifan Wang $^{1,2,4}$, Bo Ji$^{1,2,3}$, Wanpeng Gao$^{1,2,3}$,\\ Zhirong Lin$^{1,2,3}$\thanks{Corresponding author. E-mail:zrlin@mail.sim.ac.cn}, and Zhen Wang $^{1,2,3,4}$\\
$^{1}${National Key Laboratory of Materials for Integrated Circuits,} \\{Shanghai Institute of Microsystem and Information Technology, Chinese Academy of Sciences,}\\{ Shanghai 200050, China}\\  % The line break was forced via \\
$^{2}${CAS Center for Excellence in Superconducting Electronics,}\\{ 865 Changning Rd., Shanghai 200050, China}\\ % The line break was forced via \\
$^{3}${University of Chinese Academy of Sciences, Beijing 100049, China}\\
$^{4}${ShanghaiTech University, Shanghai 201210, China}}   % The line break was forced via \\

% 1. For Chinese authors, the name in Chinese characters should also be given. For example, Gang 
% 2. Please ensure that every author approves the submission of the manuscript
% 3. Abbreviations should not be used in the affiliations

\date{\today}
\maketitle

\begin{abstract}
Single-flux-quantum (SFQ) circuits have great potential in building cryogenic quantum-classical interfaces for scaling up superconducting quantum processors. SFQ-based quantum gates have been designed and realized. However, current control schemes are difficult to tune the driving strength to qubits, which restricts the gate length and usually induces leakage to unwanted levels. In this study, we design the scheme and corresponding pulse generator circuit to continuously adjust the driving strength by coupling SFQ pulses with variable intervals. This scheme not only provides a way to adjust the SFQ-based gate length, but also proposes the possibility to tune the driving strength envelope. Simulations show that our scheme can suppress leakage to unwanted levels and reduce the error of SFQ-based Clifford gates by more than an order of magnitude.
\end{abstract}

\textbf{Keywords:} superconducting qubit; qubit control; SFQ circuit

%\textbf{PACS:} no more than four \href{http://cpb.iphy.ac.cn/EN/column/item208.shtml}{PACS codes} should be provided

\section{Introduction}
Superconducting circuits offer a promising platform for constructing large-scale quantum processors \cite{doi:10.1126/science.1231930,arute2019quantum,gong2021quantum}. Currently, most superconducting quantum circuits in dilution refrigerators are controlled by shaped microwave pulses, generated by room-temperature electronics and delivered through coaxial cables. With the increasing number of qubits, such approaches become increasingly difficult due to hardware overhead, heat load management, and signal latency. For further scaling up of superconducting quantum computing, several cryogenic quantum-classical interfaces have emerged to circumvent this bottleneck, such as photonic links \cite{lecocq2021control}, cryogenic CMOS-based circuits \cite{pauka2021cryogenic,8702452}, single flux quantum (SFQ) circuits \cite{mcdermott2018quantum,8993634}, etc.  

SFQ circuit \cite{likharev1991rsfq} is a kind of superconducting digital circuit with ultra-low power consumption and excellent compatibility to superconducting qubits. By synthesizing the SFQ voltage pulses whose time integral is exactly equal to $\Phi_\mathrm{0} = \mathrm{h/2e} $, waveforms comprising of SFQ pulse trains are generated by the SFQ circuits. Universal quantum gates can be realized by coupling such specific waveforms as drive pulses to qubits, which have been demonstrated theoretically \cite{jokar2022digiq,PhysRevApplied.2.014007,PhysRevApplied.19.044031,9605311} and experimentally \cite{PhysRevApplied.11.014009,PRXQuantum.3.010350,liu2023single}. In the original control scheme, SFQ pulses distributed at equal intervals of the clock cycle constitute sequences (single-SFQ-pulse sequence) corresponding to the gate operation, which means that the driving strength remains constant during gate operation. This restricts the gate length and often induces leakage to higher levels of the qubit \cite{PhysRevApplied.2.014007}. Also, approaches including genetic algorithms were employed to optimize SFQ pulse sequences for leakage suppression \cite{PhysRevApplied.6.024022,li2019hardware}. But the generation of such sequences often requires a far larger circuit scale consisting of more than hundreds of Josephson junctions \cite{mcdermott2018quantum}. 

In this work, we propose an SFQ-based qubit control scheme with continuously tunable driving strength by dual-SFQ-pulse sequence. And we design a circuit that implements this scheme using only eleven Josephson junctions. Simulation and numerical analysis show that the circuit we designed can continuously adjust the qubit drive strength within a range of about twice that of equally spaced pulses. By tuning and shaping the driving strength, the SFQ-induced leakage to higher energy levels of the qubit is significantly suppressed. We demonstrate that this control scheme can lower the error per Clifford gate by more than a tenfold reduction.

%%%%%%%%%%%%%%%%%%%%%%%%%%%%%%%%%%%%%%%%%%
\section{The Model and Circuit Design}

\subsection{Model of Driving Strength Tuning}

\begin{figure}[t]
\centering
\includegraphics[width=16 cm]{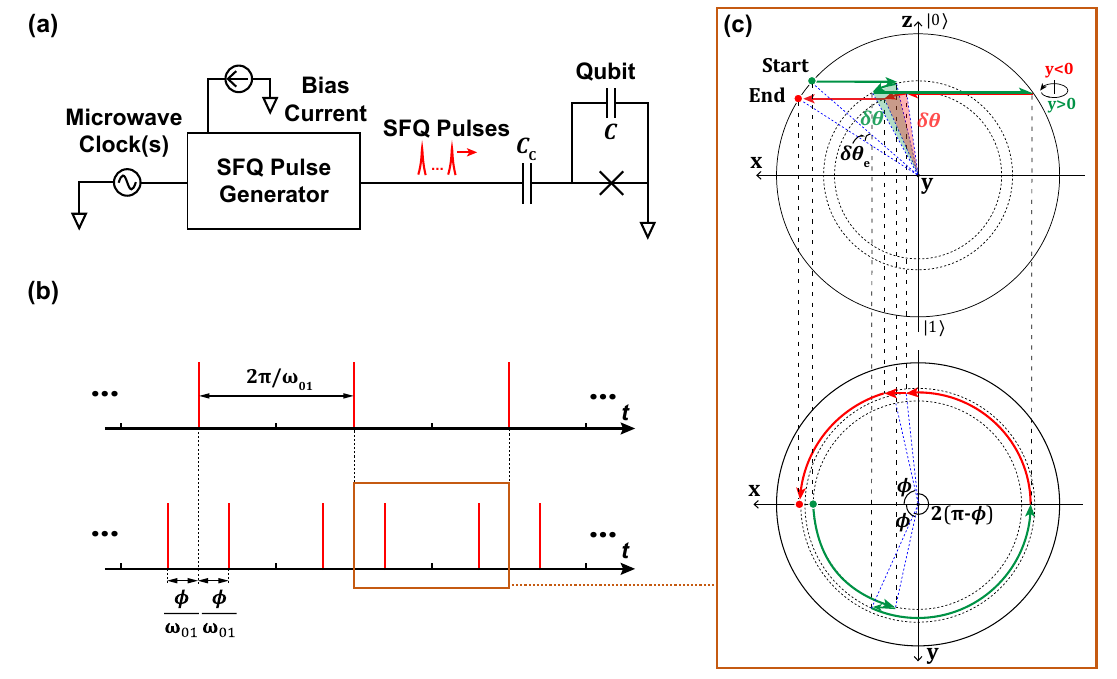}
\caption{(\textbf{a}) Simplified diagram of SFQ-based qubit control scheme. (\textbf{b}) SFQ pulse sequence. The red vertical lines depict SFQ pulses, and the  width (typically several picoseconds) of the SFQ pulse is  significantly shorter than the clock period. The upper sequence is in resonance with the qubit and the SFQ pulses are uniformly spaced. The bottom sequence corresponds to the scheme with tunable driving strength, where each pulse is replaced by a dual pulse with an adjustable interval. (\textbf{c}) The evolution trajectory of the qubit states on Bloch sphere corresponding to a clock period. The green and red arrows are the evolution trajectories in the $y>0$ and $y<0$ hemispheres, respectively.  \label{fig1}}
\end{figure}

The SFQ-based qubit control model is shown in Figure \ref{fig1}a, SFQ pulses are generated and then coupled to the transmon qubit through a capacitor $C_\mathbf{C}$. Typically, the SFQ pulses are resonant with the qubit and equally spaced, each pulse inducing a rotation of the state vector on the Bloch sphere around the y-axis by an angle $\delta\theta =C_\mathrm{C}\Phi_0\sqrt{{2\omega_{01}}/({\hbar C}}) $ \cite{PhysRevApplied.2.014007}, where $C_\mathbf{C}$ is the coupling capacitance, $\omega_{01}$ is the transition frequency of qubit, $C$ is the qubit self-capacitance. Within the two-level subspace, the qubit evolution operator including the free precession and the SFQ-pulse-induced discrete rotation of the state vector in each clock cycle can be written as $U_0=R_y(\delta\theta)R_z(2\pi)$, where $R$ is the rotation gate. In the experiment, after the device is fabricated and the qubit frequency is fixed, the incremental rotation, $\delta\theta$, of the qubit state vector in each clock cycle cannot be adjusted. 

In the control scheme of single-SFQ-pulse sequence, the phases of state vector precession are the same when each SFQ pulse triggers the qubit. To tune $\delta\theta$, the phases of the triggering moments are shifted by $\pm\,\phi$, where $0< \phi <\pi$. So, each equally spaced SFQ pulse is replaced by a double pulse with an interval $\mathrm{2}\phi/\omega_\mathrm{01}$ as shown in Figure \ref{fig1}b. The corresponding evolution operator for each clock cycle is
\begin{linenomath}
\begin{eqnarray}\label{equ1}
U&=&R_z(\phi)R_y(\delta\theta)R_z(2\pi-2\phi)R_y(\delta\theta)R_z(\phi )  \nonumber \\
~&=& 
\begin{bmatrix}
 -\cos^{2}\frac{\delta\theta}{2}-\sin^{2}\frac{\delta\theta}{2}\cdot e^{[-i(2\phi-\pi)]}   & \cos\phi\cdot\sin\delta\theta      \\
 -\cos\phi\cdot\sin\delta\theta & -\cos^{2}\frac{\delta\theta}{2}-\sin^{2}\frac{\delta\theta}{2}\cdot e^{[i(2\phi-\pi)]}
\end{bmatrix}.
\end{eqnarray}
\end{linenomath}
\unskip
In experiments, a quantum gate usually contains at least tens of pulses \cite{PhysRevApplied.11.014009,liu2023single}, that is, $\delta\theta\ll\pi/2$. So the time evolution operator can be approximated as
\begin{linenomath}
\begin{equation}\label{equ2}
U\approx R_y(2\cos\phi \cdot\delta\theta )\cdot R_z(2\pi).
\end{equation}
\end{linenomath}
The evolution trajectory of the qubit state vector within a precession period is depicted in Figure \ref{fig1}c, where the equivalent incremental rotation in each clock cycle is $\delta\theta _\mathrm{e}=2\cos\phi\cdot\delta\theta$ . By adjusting the interval of the dual pulses, we can tune the equivalent driving strength of the SFQ sequence for the qubit.

\subsection{Circuit Design}
\begin{figure}[h]
\centering
\includegraphics[width=16 cm]{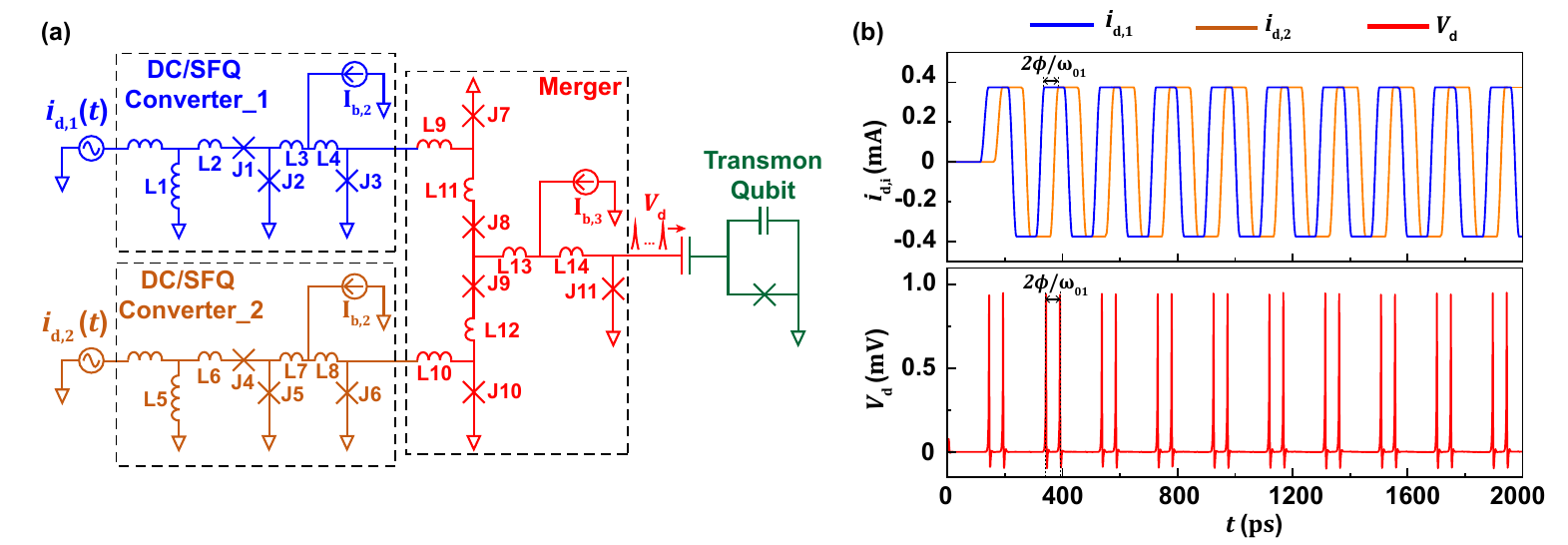}
\caption{(\textbf{a}) Circuit diagram of dual-SFQ-pulse generator driving the qubit. Two SFQ pulse sequences respectively generated by asynchronously driven DC/SFQ converters are combined into one sequence via Merger. The dual-pulse train is capacitively coupled to the transmon qubit. (\textbf{b}) Input (up) and output (bottom) in the analog simulation, where $i_\mathrm{d,i}$ is the current of microwave drive and $V_\mathrm{d}$ is the generated waveforms of the dual-SFQ-pulse generator.    \label{fig2}}
\end{figure}

We design the dual-SFQ-pulse sequence generator as shown in Figure \ref{fig2}a to realize the above control scheme. This circuit consists of two DC/SFQ converters and an SFQ merger \cite{likharev1991rsfq}. With a bias current $\mathrm{I_b}$, each DC/SFQ converter is driven by microwave pulses $i_\mathrm{d,i}$ and generates sequences of SFQ pulses corresponding to its period and phase. The SFQ merger is commonly employed as an OR gate in SFQ logic, wherein it merges the two input SFQ pulse sequences into a single output sequence.
Combining two DC/SFQ converters driven by microwaves with different phases to the merger, a dual-SFQ-pulse sequence with a specific spacing is generated. The design of the dual-SFQ-pulse sequence generator was simulated and confirmed with PSCAN2 \cite{shevchenkopscan2}.  The simulating waveforms in Figure \ref{fig2}b show that the circuit generates the dual-SFQ-pulse sequence with a spacing of $2\phi/\omega_{01}$ when the phase difference between the two microwave drives is $2\phi$. Furthermore, the circuit can operate within the range of $\phi \in (0.0423\pi,\,0.958\pi)$ with a clock frequency of 5\,GHz in margin simulation, and the limit of the range is determined by the width of the SFQ pulse.
\begin{table}[h]
\caption{Parameters of dual-SFQ-pulse generator in simulation.\label{tab1}}
%\newcolumntype{C}{>{\centering\arraybackslash}X} 
\begin{tabular}{ccllccllccccc}
\toprule[2pt]
\multicolumn{2}{l}{\textbf{DC/SFQ Converter\_1}} &  &  & \multicolumn{2}{l}{\textbf{DC/SFQ Converter\_2}} &  &  & \multicolumn{5}{c}{\textbf{Merger}}                                  \\ \hline
J1                    & 337 ${\mu}$A               &  &  & J4                    & 337 ${\mu}$A               &  &  & J7     & 262 ${\mu}$A &  & L9                   & 3.3 pH                 \\
J2                    & 180 ${\mu}$A               &  &  & J5                    & 180 ${\mu}$A               &  &  & J8     & 237 ${\mu}$A &  & L10                  & 3.3 pH                 \\
J3                    & 235 ${\mu}$A               &  &  & J6                    & 235 ${\mu}$A               &  &  & J9     & 237 ${\mu}$A &  & L11                  & 1.15 pH                 \\
L1                    & 4.45 pH                     &  &  & L5                    & 4.45 pH                     &  &  & J10    & 262 ${\mu}$A &  & L12                  & 1.15 pH                 \\
L2                    & 1.05 pH                     &  &  & L6                    & 1.05 pH                     &  &  & J11    & 312 ${\mu}$A &  & L13                  & 1.25 pH                 \\
L3                    & 2.35 pH                     &  &  & L7                    & 2.35 pH                     &  &  & $\mathrm{I_{b,3}}$ & 487 ${\mu}$A &  & L14                  & 4.2 pH                 \\
L4                    & 2.65 pH                     &  &  & L8                    & 2.65 pH                     &  &  &        &            &  &                      & \multicolumn{1}{l}{} \\
$\mathrm{I_{b,1}}$                & 350 ${\mu}$A               &  &  & $\mathrm{I_{b,2}}$                & 350 ${\mu}$A               &  &  &        &            &  & \multicolumn{1}{l}{} & \multicolumn{1}{l}{} \\ 
%\hline
\bottomrule[2pt]
\end{tabular}
\end{table}

%%%%%%%%%%%%%%%%%%%%%%%%%%%%%%%%%%%%%%%%%%
\section{Simulation and Analysis}
\subsection{Driving Strength Spectrum Analysis}

\begin{figure}[h]
\centering
\includegraphics[width=16 cm]{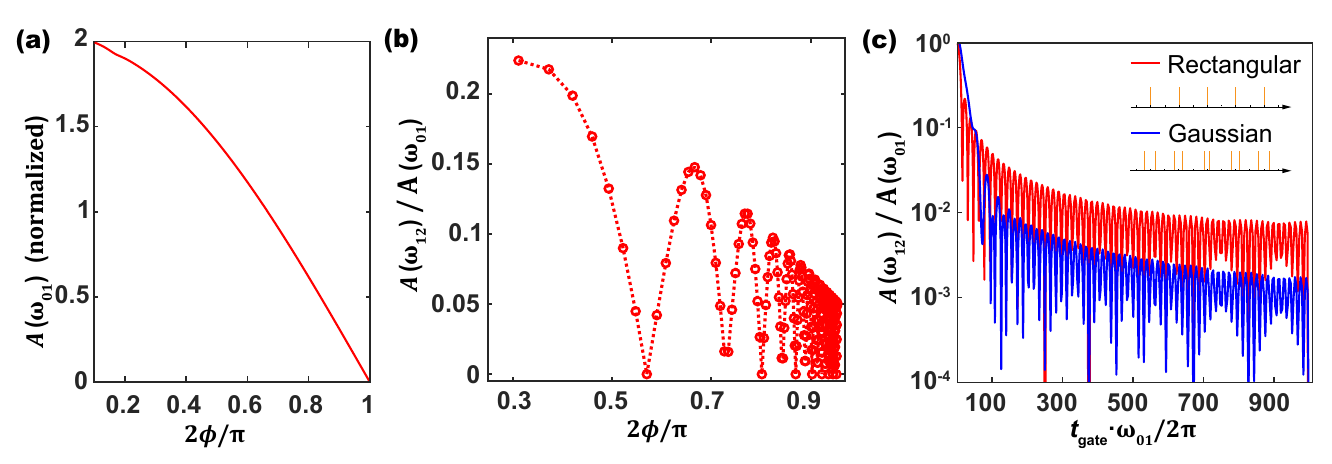}
\caption{Spectrum analysis of dual-SFQ-pulse sequences for $\omega_{01}/2\pi=5\; \mathrm{GHz}$ and $\omega_{12}/2\pi=4.6\,\mathrm{GHz}$. (\textbf{a}) The driving signal spectral component ${A} (\omega _{01})$ at the resonant frequency as a function of the phase difference $2\phi$ between $i_\mathrm{d,i}$. (\textbf{b}) The ratio ${A} (\omega _{12})/{A} (\omega _{01})$ of a $\pi$ pulse as a function of $2\phi$ with $\delta \theta=\pi/30$. (\textbf{c}) The ${A} (\omega _{12})/{A} (\omega _{01})$ of sequences with rectangle-like and Gaussian-like driving strength envelope as a function of gate length $t_\mathrm{gate}$.   \label{fig3}}
\end{figure}

We apply the Fourier transform on the voltage waveform of the dual-SFQ-pulse sequence generated in the ciucuit simulation. As shown in Figure \ref{fig3}a, the spectral component of driving sequence at the resonant frequency ${A} (\omega_{01} )$ can be tuned across 0 to 1.98 times the value in the equally spaced sequence by modulating $\phi$, which agrees with the incremental rotation of the qubit state vector $\delta\theta _\mathrm{e}$. While it may appear that extending the length of gate operation by ${A} (\omega_{01} )$ tuning can attenuate the spectral component at the frequency of the unwanted transition (such as 1-2 transition), it is should be noted that the spectral component corresponding to the 0-1 transition is also reduced \cite{li2019hardware}. To evaluate the leakage to higher energy levels in qubits driven by dual-SFQ-pulse sequences, we compare the spectra of the sequences for the same gate operation. Upon increasing $\phi$, the ratio between spectral components at $\omega_{12}$ and $\omega_{01}$  shows a gradual decreasing trend with fluctuations, as depicted in Figure \ref{fig3}b. This suggests that appropriate modulation of $\phi$ can effectively suppress leakage to higher energy levels.

The driving strength of an SFQ pulse train to the qubit depends on its spectral component at the resonant frequency ${A} (\omega _{01} )$. The ${A} (\omega _{01} )$ is fixed during the single-SFQ-pulse sequence, resulting in a rectangle-like driving strength envelope. In microwave-based control schemes, the drive strength envelope is usually optimized to minimize leakage to unwanted energy levels \cite{10.1063/1.5089550,PhysRevB.68.224518}. The dual-SFQ-pulse sequence now offers the potential to optimize the envelope for SFQ-based qubit control by tuning the dual-pulse interval per clock cycle. For instance, as demonstrated in Figure \ref{fig3}c, when the driving strength envelope is modulated into a Gaussian-like shape, we observe a faster decrease of the ratio ${A} (\omega _{12})/{A} (\omega _{01})$ with the gate length compared to the single-SFQ-pulse sequence with rectangle-like envelope. However, it is worth noting that accurate regulation of the driving strength envelope depends on the accurately regulated two-pulse intervals per clock cycle, which means that arbitrary waveform generators with sampling rates much higher than the clock frequency are required to generate $i_\mathrm{d,i}$ in the circuit in Figure \ref{fig2}a.

\subsection{Fidelity Optimization}
Quantum gates implemented using dual-pulse sequences are expected to exhibit higher fidelity compared to single-pulse sequences with equally spaced SFQ pulses, due to the suppression of leakage to non-computational states. To quantitatively assess the potential of the dual-SFQ-pulse driving scheme in enhancing fidelity, we conducted randomized benchmarking with simulation to evaluate the fidelity of Clifford gates based on single- and dual-SFQ pulse sequences. Initially, we modeled the transmon as a three-level system, with the output voltage waveform $V(t)$ from the SFQ pulse generator coupled to it, as depicted in Hamiltonian
\begin{linenomath}
\begin{equation}\label{equ3}
\hat{H}= \hbar\left[\begin{array}{ccc}
0 & 0 & 0 \\
0 & \omega_{01} & 0 \\
0 & 0 &  2\omega_{01}- \alpha 
\end{array}\right]+i\frac{\hbar\delta \theta }{\Phi _0} V(t)\left[\begin{array}{ccc}
0 & -1 & 0 \\
1 & 0 & -\sqrt{2} \\
0 & \sqrt{2} & 0
\end{array}\right].
\end{equation}
\end{linenomath}
We then calculated the qubit state evolution and sequence visibility numerically by QuTiP \cite{JOHANSSON20131234}. This allowed us to calibrate the pulse sequences for each quantum gate and perform randomized benchmarking for parameters of the transmon qubit and coupler listed in Table \ref{tab2}.

\begin{table}[H] 
\centering
\caption{Parameters of randomized benchmarking.\label{tab2}}
%\newcolumntype{C}{>{\centering\arraybackslash}X}
\begin{tabularx}{\textwidth}{cccc}
\toprule[1pt]
\makebox[0.2\textwidth][c]{Parameter}	& \makebox[0.2\textwidth][c]{$\delta \theta$}	&\makebox[0.2\textwidth][c]{$\omega_{01}/2\pi$}	& \makebox[0.3\textwidth][c]{Anharmonicity $\left |\alpha   \right |/2\pi$}\\
\midrule
I		& $\pi/30$        & 5 GHz         & 400 MHz \\
II		& $\pi/60$        & 5 GHz         & 450 MHz \\
\bottomrule
\end{tabularx}
\end{table}

\begin{figure}[h]
\centering
\includegraphics[width=14 cm]{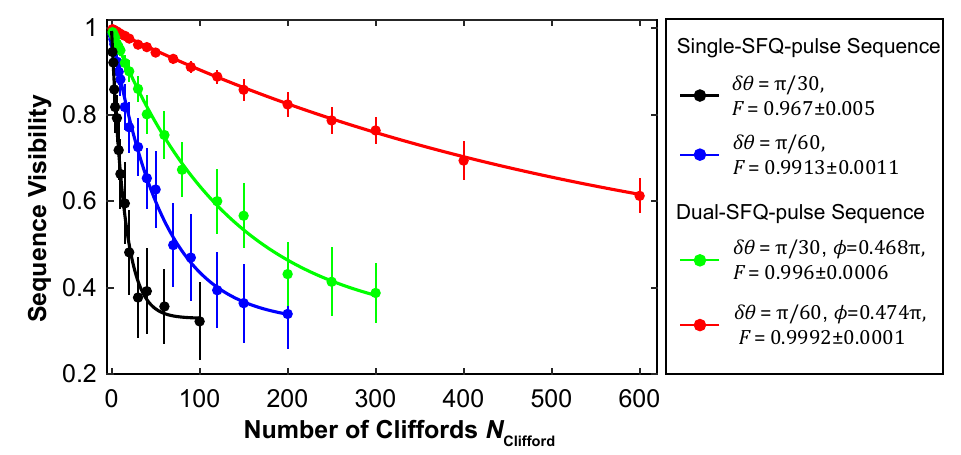}
\caption{Randomized Benchmarking of SFQ-based gates. We simulate 100 random sequences for each sequence length $N_\mathrm{Clifford}$.  \label{fig4}}
\end{figure}

As demonstrated in Figure \ref{fig4}, substituting the single-SFQ-pulse sequence with an optimized dual-SFQ-pulse sequence can considerably enhance fidelity. Even in parameter conditions where excessive coupling induces significant leakage ( $\delta \theta=\pi/30 $ ), the average fidelity of Clifford gates is improved from 96.7 $\%$ to 99.6 $\%$ by tuning the driving strength with dual-SFQ-pulse sequence. Likewise, for parameter-II as shown in Table \ref{tab2}, the optimized dual-SFQ-pulse sequence suppresses the error per Clifford gate to $8\times 10^{-4}$, which is more than one order of magnitude lower.

%%%%%%%%%%%%%%%%%%%%%%%%%%%%%%%%%%%%%%%%%%
\section{Summary and Outlook}

In summary, we proposed a qubit control scheme based on SFQ circuitry with tunable driving strength and present a dual-SFQ-pulse sequence generator to implement it. Through simulation and analysis, we demonstrate that the pulse generator can produce sequences with continuously tunable driving strength and fixed coupling. This overcomes the limitation of the original single-SFQ-pulse sequence, which lacked the ability to shape the driving strength envelope. By tuning and shaping the driving strength, the power spectral component inducing leakage to the non-computational state can be significantly suppressed. The randomized benchmarking results show that compared with the Clifford gate composed of the single-SFQ-pulse sequence, the optimized dual-SFQ-pulse sequence reduces the gate error by more than an order of magnitude.

Moreover, our proposed qubit control scheme with tunable driving strength not only improves the fidelity of SFQ-based single-bit gates but also has practical significance in SFQ-Activated CZ gate \cite{PhysRevApplied.19.044031}. It provides another dimension of tunability and allows the two-qubit gate scheme to arbitrarily select working points with different gate times under specific subharmonic. This optimization makes the SFQ-Activated CZ gate scheme more suitable for all-fixed qubits systems.

%%%%%%%%%%%%%%%%%%%%%%%%%%%%%%%%%%%%%%%%%%
\vspace{6pt} 

%%%%%%%%%%%%%%%%%%%%%%%%%%%%%%%%%%%%%%%%%%
%% optional
%\supplementary{The following supporting information can be downloaded at:  \linksupplementary{s1}, Figure S1: title; Table S1: title; Video S1: title.}


%%%%%%%%%%%%%%%%%%%%%%%%%%%%%%%%%%%%%%%%%%

%\institutionalreview{}
%\dataavailability{} 

%\acknowledgments

\noindent \textbf{Acknowledgment}s:  

{This work is supported in part by the National Natural Science Foundation of China (No.92065116), the Key-Area Research and Development Program of Guangdong Province, China (No.2020B0303030002), the Shanghai Technology Innovation Action Plan Integrated Circuit Technology Support Program (No.22DZ1100200) and Strategic Priority Research Program of the Chinese Academy of Sciences (Grant No.XDA18000000). We thank Dr.\;Jie\;Ren, from Shanghai Institute of Microsystem and Information Technology (SIMIT), CAS, for providing SFQ design infrastructure.\;}

\end{CJK*}  %% end the Chinese environment
\end{document}